\documentstyle[epsf]{mn2e}

\title[Spin determination from disc spectra in GX 339$-$4]
{Limits on spin determination from disc spectral fitting in GX 339$-$4}

\author[M. Kolehmainen \& C. Done]
{Mari Kolehmainen $^1\thanks{E-mail:m.j.kolehmainen@durham.ac.uk}$ and Chris Done $^1$ \\ 
$^1$Department of Physics, University of Durham, South Road, Durham
DH1 3LE,
UK\\
}

\pagerange{\pageref{firstpage}--\pageref{lastpage}} \pubyear{2009}

\begin{document}

\topmargin = -0.5cm

\maketitle

\label{firstpage}

\begin{abstract}

We attempt to constrain the black hole spin in GX 339$-$4 from
spectral fitting of disc dominated data using \textsl{RXTE} spectra
from the three most recent outbursts.  We use the best current models
for the disc emission, including full radiative transfer through the
photosphere rather than assuming that the intrinsic emission from each
radius has a (colour temperature corrected) blackbody spectrum. The
results strongly depend on the poorly known binary system parameters,
but we find a strict upper limit of $a_{*}<0.9$ for any distance
greater than 6~kpc, assuming that the orbital inclination is the same
as that of the inner disc. By contrast, the higher spin of 0.935 +/-
0.01 (statistical) +/-0.01 (systematic) claimed from fitting the iron
line profile in this object requires that the inner disc is misaligned
by over 20 degrees from the orbital inclination.  While some of these
datasets are distorted by instrumental pileup, the same
spin/inclination constraints are derived from data which are not piled
up, so there is a real conflict between the two techniques to measure
spin. We argue that the disc spectral fits are more likely to be
robust hence that there are still issues to be understood in the iron
line profile.

\end{abstract}

\section{Introduction}

An astrophysical black hole (BH) can be entirely described by two
parameters in general relativity, its mass \textsl{M} and a
dimensionless spin, $a_{*}$, which is 0 for a non-rotating
Schwarzschild BH and $0.998$ for a maximal Kerr BH.  Unlike mass, spin
only leaves an imprint on the spacetime very close to the event
horizon, so it is much more difficult to measure.  Nonetheless, it is
important to constrain because it is a fundamental parameter
determining the structure of the spacetime around the BH. It
sets the size--scale of the last stable orbit around the BH, from
$6-1.23R_g$ for $a_*=0$ and $0.998$, respectively, where
$R_g=GM/c^2$. This determines the efficiency of conversion of mass to
radiation for accreting objects, and may also determine the
structure and power of relativistic jets. For BHs formed from stellar
collapse then the resulting spin gives insight into the (poorly
understood) supernovae event (Gammie, Shapiro \& McKinney 2004) and
its gravitational wave signature.

Currently there are only two methods which can be used to determine
spin from accreting BHs (see Reynolds \& Fabian 2008; McClintock \&
Remillard 2006). The first uses the luminosity and temperature of the
optically thick, geometrically thin accretion disc to measure the
emitting area of the inner disc, and hence its radius. This assumes
that the dissipation follows that of the relativistic stress-free
inner boundary condition (Novikov \& Thorne 1973), an assumption which
is now strongly supported by recent fully relativistic MHD simulations
of thin discs with self-consistent magnetic turbulence as the origin
of stress (Shafee et al. 2008).

To use this method requires that we can observe the spectrum at
energies close to the peak temperature, which limits this method to
stellar mass BH binaries (hereafter BHBs), as AGN discs typically peak
in the unobservable far UV.  Additionally, the source distance, and
disc inclination (assumed to be the same as that of the binary) must
be constrained, as these are necessary to transform the observed disc
flux into luminosity. Similarly, the BH mass is required to convert
the resulting size scale into gravitational radii. Thus this technique
can only be used on a small subset of systems for which this
information is available. Further restrictions are that the systems
should be {\em dominated} by the disc emission. The method can still
be applied to data with an increasing fraction of emission in the
power law tail, but with increasingly large uncertainties in
reconstructing the temperature and luminosity of the disc emission
(Kubota et al. 2001; Kubota \& Done 2004; Done \& Kubota 2006; Steiner
et al. 2009). A final restriction is that the source should not be
too bright as the disc structure may change at luminosities approaching and
exceeding the Eddington limit $L_{\rm{Edd}}$. The disc can puff up to
become geometrically thick, advection and winds may become important,
and the spectrum may also be increasingly distorted by low
temperature Comptonization which can be difficult to distinguish from
disc emission (e.g. GRS 1915+105: where 
Middleton et al. 2006 derive $a_*\sim 0.7$ 
compared to $a_*=0.98$ from McClintock et al. 2006 due to differences in 
Comptonisation assumptions).

Within these limitations, the models including the full physics
(stress-free inner boundary condition, relativistic smearing, and
modelling the non-blackbody intrinsic emission from each radial
annulus) are remarkably robust to changes in the disc vertical
structure from different {\em ad hoc} stress prescriptions (Done \&
Davis 2008), and all the potential uncertainties act in the same
direction which is for these models to overestimate the black
hole spin (Gierli\'nski \& Done 2004; Done \& Davis 2008).

The second method to measure spin uses the shape of the iron line
produced by fluorescence in the X-ray illuminated accretion disc.  The
width of the line is set by the line emissivity, together with
the strength of the gravitational field as this determines the
velocity of the disc (hence the Doppler shift, beaming and time
dilation) as well as gravitational redshift. All these parameters
(inclination, inner disc radius in terms of gravitational radii and
emissivity) can be constrained directly from spectral
fitting, so this technique can be used much more widely than disc
spectral fitting. Its only restrictions are that the disc is flat and
in Keplarian rotation (which again becomes increasingly uncertain at
luminosities approaching/exceeding Eddington) and that there are
sufficient hard X-rays illuminating the disc to produce the iron line.
Thus it can be used for both (sub-Eddington) AGNs and BHBs and requires
no additional information about the distance and/or inclination of the
system (Fabian et al. 1989; Fabian et al. 2000).

However, unlike the disc in the disc dominated spectra, the line is
only a small feature on the total spectrum, which means it can be
difficult to measure. The line sits on top of a reflected continuum,
and the shape of both line and reflected continuum depend on
ionisation of the material (Ross, Fabian \& Young 2001), and the
radial and vertical profile of this ionisation (Nayakshin, Kazanzas \&
Kallman 2001; Done \& Nayakshin 2007). Other key issues are the
underlying continuum shape (which can be distorted by complex
absorption: L. Miller et al. 2007; 2008) and disentangling the intrinsic
shape of the blue wing of the line from any absorption lines from
ionised iron K$\alpha$ (Done \& Gierli\'nski 2006; Done \& Kubota
2006; Young et al. 2005).

Since we have two methods to measure spin it is obviously important to
compare them. We would have increased confidence in both methods if
they gave the same answer for the same object. However, the very
different restrictions on the two techniques mean there is a very
small set of objects where both can be used.  Disc dominated spectra
generally have too few photons at the high energies required to
produce a strong iron fluorescence line. Similarly, disc spectral
fitting cannot be used on spectra with a strong tail (carrying more
than 25 per cent of the bolometric luminosity) as the uncertainties in
reconstructing the intrinsic disc emission become too large (Kubota \&
Done 2004).  Thus the
two methods cannot be reliably compared using the same dataset (but
see Miller et al. 2009 for an attempt at this), but they can be used on
the same object for a BHB with well constrained system parameters
which shows spectral transitions.

To date only three objects have good spin estimates from both methods,
where 'good' is defined as derived from fits with the best currently
available models i.e. {\sc bhspec} for disc spectral fitting (Davis et
al. 2005) and {\sc CDID} for ionised reflection (Ross \& Fabian 2005).
The results are not encouraging. 
The spin estimates match well only for XTE~J1550-564, are somewhat
discrepant for 4U~1543-475, and are quite significantly different for
GRO~J1655-40 (see Section~\ref{s:compare}).

It is clearly important to expand the sample of objects for which this
comparison can be made. GX 339$-$4 is one of the best studied BHBs in
terms of iron line profile from three separate data sets, spanning a
range of spectral states (Miller et al. 2004; 2006; 2008; Reis et al.
2008). Two of these datasets have issues with pileup, which
distorts the line (Done \& Diaz-Trigo 2010; Yamada et al.
2010), but results from the third dataset alone indicate a very high
spin, with $r_{in}\sim 1.8$ or $2$ depending on the detailed emissivity
profile, requiring $a_*=0.96$ or $0.935$ (Reis et al. 2008) and 
an inclination of $\sim 20^\circ$. 

Here we apply the disc spectral fitting method and find that, despite
the poorly known system parameters, such high spin is unlikely to
match the observed temperature and luminosity of the disc dominated
spectra from this source. 

\section{System parameters}

Measuring spin from disc spectral fitting requires that the system
parameters (mass, distance and inclination) are reasonably
constrained. Without relativistic effects, the disc luminosity relates
to the inclination via $L\propto 2{\pi}D^{2} F / {\rm{cos}} i\propto A
T_{\rm{disc}}^4$, where D is distance, F is the observed flux and
$A=2\pi r_{in}^2 R_g^2$ is the area of the inner disc. Thus
$r_{in}^2\propto D^{2}/(M^2\cos i)$. Increasing $D/M$ and increasing
inclination means larger $r_{in}$ and hence smaller spin. 

The binary parameters are more easily determined in X-ray quiescence,
when the compact object is faint so that the secondary star can be
seen. The spectral type then gives distance, while the ellipsoidal
variations give the mass and inclination. Alternatively, the mass and
inclination can also be constrained from outburst data, by tracking
periodic shifts in the line emission from the X-ray irradiated star
(Casares 2007). 

For GX 339$-$4, the continuum from the secondary star cannot be
clearly seen even during quiescence but detection of the line emission
in outburst gives a mass function of 5.8 $\pm$ 0.5 $M_{\odot}$ (Hynes
et al. 2003, hereafter H03). We review constraints on the mass,
distance and inclination below, and then consider a representative
selection of these values in order to explore their impact on the
derived BH spin.

\subsection{Distance, mass and inclination}

Since the distance cannot be constrained by the companion star, it is
instead estimated from the Na D absorption along the line of sight to
the source. This gives $D\geq 6$~kpc, with distances as large as 15~kpc
allowed, which would place GX 339$-$4 on the far side of the Galaxy
(Hynes et al. 2004). Zdziarski et al. (2004) (hereafter Z04), show
that the Na D absorption to GX 339$-$4 is similar to that seen towards
OB stars in the galactic bulge, so suggest that it is more likely that
GX 339$-$4 is at $D=8$~kpc. Thus we take $6<D<15$~kpc as our range in
distance, but this is probably too conservative as the companion star
is likely to be seen at a distance of 6~kpc (Mu\~noz-Darias,
Casares \& Martinez-Pais 2008)

The mass function gives a lower limit to the mass of $5.3$ $M_{\odot}$
(H03). However, this is too conservative considering the companion
star. This must fill its Roche lobe, which is only possible with a
long orbital period for a somewhat evolved star.  Mu\~noz-Darias,
et al. (2008) look in detail at these constraints,
and argue for a stripped giant companion star with mass $M_{2} \geq
0.166$ $M_{\odot}$, giving a solid lower limit to the black hole mass of
$6.2~M_\odot$. Conversely, the largest mass black hole in a low mass
X-ray binary is GRS 1915+105 at $14\pm 1M_\odot$. Thus we consider the
range $5.8-15$ $M_\odot$ to be conservative, but note that $6.2-15$
 $M_\odot$ is more likely.

There is a strong constraint from the lack of eclipses that
$i<80^\circ$ (H03). However, the long orbital period and consequent
large disc mean that this source almost certainly has a strong
equatorial disc wind in its high luminosity states. These give strong
ionised absorption lines when viewed at high inclinations
($i>70^\circ$). There are ionised absorption features seen in GX 339$-$4,
but these are much weaker than seen in the high inclination objects
(Miller et al. 2004). This argues for an intermediate outer disc
inclination, somewhat less than $70^\circ$, but not so much lower that
the equatorial wind does not intercept the line of sight. Thus the
outer disc (which should have the same inclination as the binary
orbit) must have an inclination $\approx 50^\circ-70^\circ$.  The
upper limit of this range is similar to the hard limit on the orbital
inclination of $i>45^\circ$ that comes from putting the maximum BH
mass of 15 $M_\odot$ into the H03 mass function with the minimum
companion star mass of $0.166$ (Mu\~noz-Darias et al. 2008).

However, it is the inner disc inclination which is important for
determining spin, and this can be misaligned from the binary/outer
disc if the BH spin is misaligned. There is a weak observational
constraint on the inner disc inclination of $i<70^\circ$ from the fact
that there are no high frequency QPOs detected in this source
(Schnittman, Miller \& Homan 2006), and a strong requirement from the
iron line fits at $i\sim 20^\circ$ (Miller et al. 2008). However,
this would require a large misalignment angle of more than $25^\circ$
between the BH spin and orbit. Such large misalignments can only be
produced from a very asymmetric supernovae, and the resultant large
kick is likely to unbind the orbit (Fragos et al. 2010).  Thus while we
consider the range $20^\circ<i<70^\circ$, the more likely lower limit
is $45^\circ$ from the small misalignments required to form the binary
(Fragos et al. 2010).

\subsection{Parameter sets}

From the ranges given above, we select some example system parameter
sets in order to illustrate the impact on derived spin.
Previous work on GX 339$-$4 by Gierli\'nski \& Newton (2006) argued for 10 
$M_\odot$, $D=6$~kpc and $i=40^\circ$ (hereafter termed GN06), as this
gave similar transition properties to other BHBs.  Zdziarski et al.
(2004) used 10 $M_\odot$, $D=8$~kpc and $i=60^\circ$ (hereafter termed
Z04). The highest spin values will be found from the lowest $D/M$ so
$M=15$ $M_\odot$, $D=6$~kpc and lowest inclination, set to the more
likely limit of $45^\circ$ (hereafter termed Max). However, we also
consider a parameter set with $i=20^\circ$ for the minimum BH mass and
distance of 5.8 $M_\odot$, $D=6$~kpc (hereafter termed H03).

\begin{table}
\begin{tabular}{lc|l|l||l|l|l|lc}
  \hline
  \textsl{M} ($M_{\odot}$) & &  \textsl{d} (kpc) & & \textsl{i} (deg) &  &  \\
  \hline
 5.8 & & 6 &  & $20$ &  & H03 \\
 10 &  & 8 &  & $60$  & & Z04 \\
 10 &  & 6 &  & $40$  & & GN06 \\
 15 &  & 6 &  & $45$  & & Max \\
\hline
\end{tabular}
\caption{The parameter sets used in this paper. The abbreviations refer to papers referenced in the text and are explained in Section 2. 
The inclinations are mostly indicative, chosen to allow a wider limit range.}
\end{table}

\section{Data analysis and selection of disc dominated spectra}

GX 339$-$4 has been widely observed with NASA's \textsl{RXTE}
satellite since its launch in 1995, and shows multiple dramatic
outbursts covering all spectral states (Remillard \& McClintock 2006;
Done, Gierli\'nski \& Kubota 2007, hereafter DGK07). We use data from the last three
outbursts of GX 339$-$4, namely the 2002/2003, 2004 and 2007
outbursts, where hard to soft state transitions could be observed.
Data reduction was done using the standard \textsl{RXTE} data analysis
methods. We add a systematic uncertainty of 1\% to all the PCA
spectra, and fit from 3--20 keV.

We follow the approach of Done \& Gierli\'nski (2003) in fitting the
spectra with a simple multicolour disc ({\sc diskbb}) plus thermal
Comptonization ({\sc thcomp}), with a Gaussian line ({\sc gau}, constrained
in energy between 6--7~keV) and smeared absorption edge ({\sc smedge},
constrained in energy between 7--9~keV) added to approximate the
effects of reflection. The hydrogen column density was fixed at
$N_{\rm{H}}= 6\times 10^{21} {\rm{cm}}^{-2}$ (Zdziarski et
al. 2004). We use these results to make hardness-intensity diagrams,
and select only disc dominated spectra for spectral fitting (defined
as those with HR $\le$ 0.2, which corresponds to spectra where the
disc contains more than 80 \% of the total luminosity). 

\section{Simple disc models}

\begin{figure}
\begin{center}
\leavevmode\epsfxsize=6cm\epsfbox{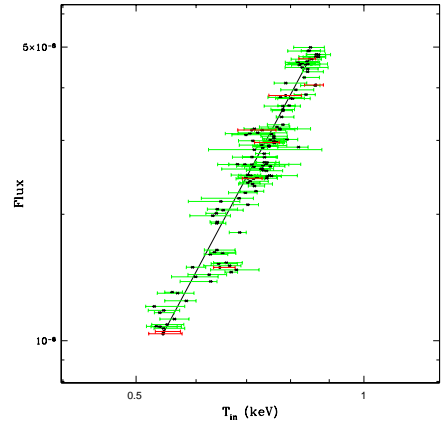}
\end{center}
\caption{The bolometric disc flux versus the inner disc temperature from
  fitting {\sc diskbb} to the disc dominated spectra from GX 339$-$4. The solid line illustrates the $T_{\rm{disc}}^4$ relation and the red dots indicate the 9 spectra chosen for simultaneous fitting in Section 5.1.
}
\label{fluxt}
\end{figure}

Within the limitations of disc spectral fitting discussed in the introduction, 
an indication of the black hole spin can be derived from fitting the very simplest disc model, {\sc diskbb} 
(Mitsuda et al. 1984), to the data. To illustrate that these disc dominated spectra not only look like a
disc, but vary like one too, we first use the simple {\sc diskbb} fits above and plot bolometric,
unabsorbed disc flux from the models against temperature in Fig~\ref{fluxt}. 
Plainly the data are consistent with $L_{\rm{disc}}\propto T_{\rm{disc}}^4$
relation, as expected from a constant inner size scale set by the last
stable orbit. This observation gives a foundation for the more physical disc models described in Section 5. 

To use these data to derive the size of this inner radius is more
challenging than simply converting flux to luminosity via the system
parameters discussed in Section 2. The derived luminosity is also 
dependent on inclination through relativistic
effects, with Doppler boosting amplifying the observed flux at high
inclinations. These relativistic effects also change the observed
temperature with Doppler blueshift dominating at high inclination
while redshifts from time dilation and strong gravity dominate at
lower inclinations. The effect on both temperature and luminosity is
small at $i=60^\circ$ for $a_*=0$, but the much stronger relativistic
effects at $a_*=0.998$ mean that red and blueshifts approximately cancel
at an inclination of $i=75^\circ$ (Zhang et al. 1997).

\begin{figure}
\begin{center}
\leavevmode \epsfxsize=8cm \epsfbox{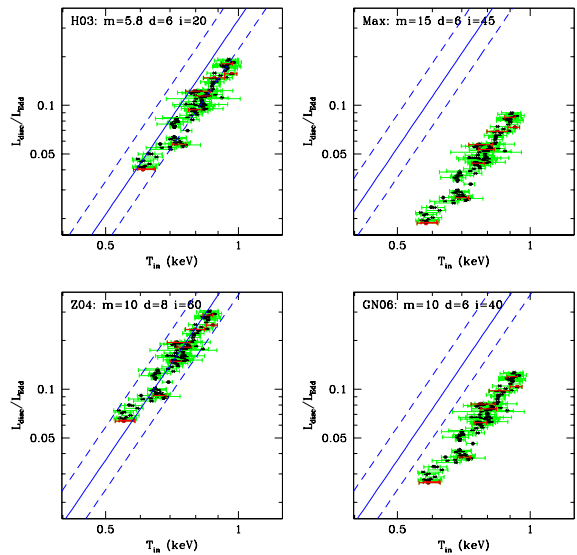}
\end{center}
\caption{The $L-T^{4}$ relation in GX 339$-$4 with the different system parameter sets detailed in Table 1. The lines refer to $f_{\rm{col}}$ values of 1.6, 1.8 and 2.0 for $a_{*}=0$. The solid line marks the best estimate of $f_{\rm{col}}=1.8$.}
\label{tlumin}
\end{figure}

We follow Gierli\'nski \& Done (2004) and correct the {\em data} for
these relativistic distortions, to derive an 'intrinsic' temperature
and luminosity for each parameter set assuming $a_*=0$.  We use the
correction factors tabulated by Zhang et al. (1997), interpolated in
$\cos i$ using a cubic. Figure~\ref{tlumin} shows these corrected data
for $a_*=0$.

Even though the simple disc models are widely used as approximations,
the true inner disc is much more complicated. The {\sc diskbb} models
assumes a temperature distribution $T(r) \propto r^{-3/4}$, so does
not incorporate the relativistic stress-free inner boundary
condition. Similarly, each annulus of the disc does not emit a true
blackbody. The continuum (free-free) absorption drops as a function of
frequency, which means that the higher energy photons in each radius
are unlikely to thermalise. The emission instead can be described as a
modified blackbody, characterised by a colour temperature which is a
factor $f_{\rm{col}}=1.6-2$ (Shimura \& Takahara 1995) above the
blackbody emission.  We incorporate both these correction factors in
the lines overlaid on Fig~\ref{tlumin}. (see Gierli\'nski \& Done 2004), for the
expected size scale of $6R_g$ for $a_*=0$. These show that the derived
spin will be dependent on the assumed system parameters. The system
parameters of Z04 appear consistent with zero spin, while H03, GN06
and Max are progressively further from the $a_*=0$ prediction,
indicating higher spins.

\section{A more physical model for disc}

Even after all the corrections to the disc spectrum, the assumption
that the spectrum has a modified blackbody shape is still just an
approximation to the full radiative transfer scenario. Photo-electric
(bound-free) absorption from partially ionised metals becomes
significant especially at high energy frequencies, where the free-free
absorption drops. This effect is highlighted by the radiative transfer
through the vertical structure of the disc, and while these spectral
features are smeared by the special and general relativistic effect,
they still result in a broader spectrum than predicted by the simpler
models. We use the {\sc bhspec} model by Davis et al. (2005)
which includes stellar atmospheres-like calculations to calculate the
vertical structure and radiative transfer of disc annuli, and incorporates
the self-consistent relativistic radial dissipation profile (stress-free
inner boundary condition for any $a_*$: Novikov \& Thorne 1976) and fully
relativistic transport to produce the observed spectrum. 

This model is described by the physical parameters of mass, spin,
distance, inclination and mass accretion rate, parameterised as
$L/L_{\rm{Edd}}$ (so the
corresponding mass accretion rate depends on the black hole spin).
Thus there is no effective temperature to use as input into
the Comptonization for the seed photon energy. Instead we fix this at
the best fit inner disc temperature derived from the previous {\sc diskbb}
models.

The large number of datasets means that simultaneous fitting for $a_*$
across all the spectra is not feasible. Instead, we fit each spectrum
separately with $a_*$ fixed at 0, then average all the individual
$\chi^2$. We then repeat this for $a_*$ fixed at 0.1, 0.2 etc up to
0.998. Fig~\ref{9spec} shows these ${\chi}^2$ versus spin for the four
parameter sets. The plots show a rough
position for the minimum $\chi^2$ i.e. best fit spin for each
particular set of parameters. As is quite clear from
Figure~\ref{9spec}, the best-fit values for $a_{*}$ vary
significantly with the different parameter sets. Z04 gives a best fit
of $a_*=0$ as indicated by the simple {\sc diskbb} fits above, while
H03, GN06 and Max have a best fit of $\sim 0.7$, $0.8$ and $0.9$,
respectively. All these best fits have very similar 
${\chi}^2$, so these very different spin values obtained from the very
different parameter sets cannot be distinguished by spectral fitting.

\subsection{Simultaneous spectral fitting }

\begin{figure*}
\begin{tabular}{cl}
\leavevmode\epsfxsize=9cm\epsfbox{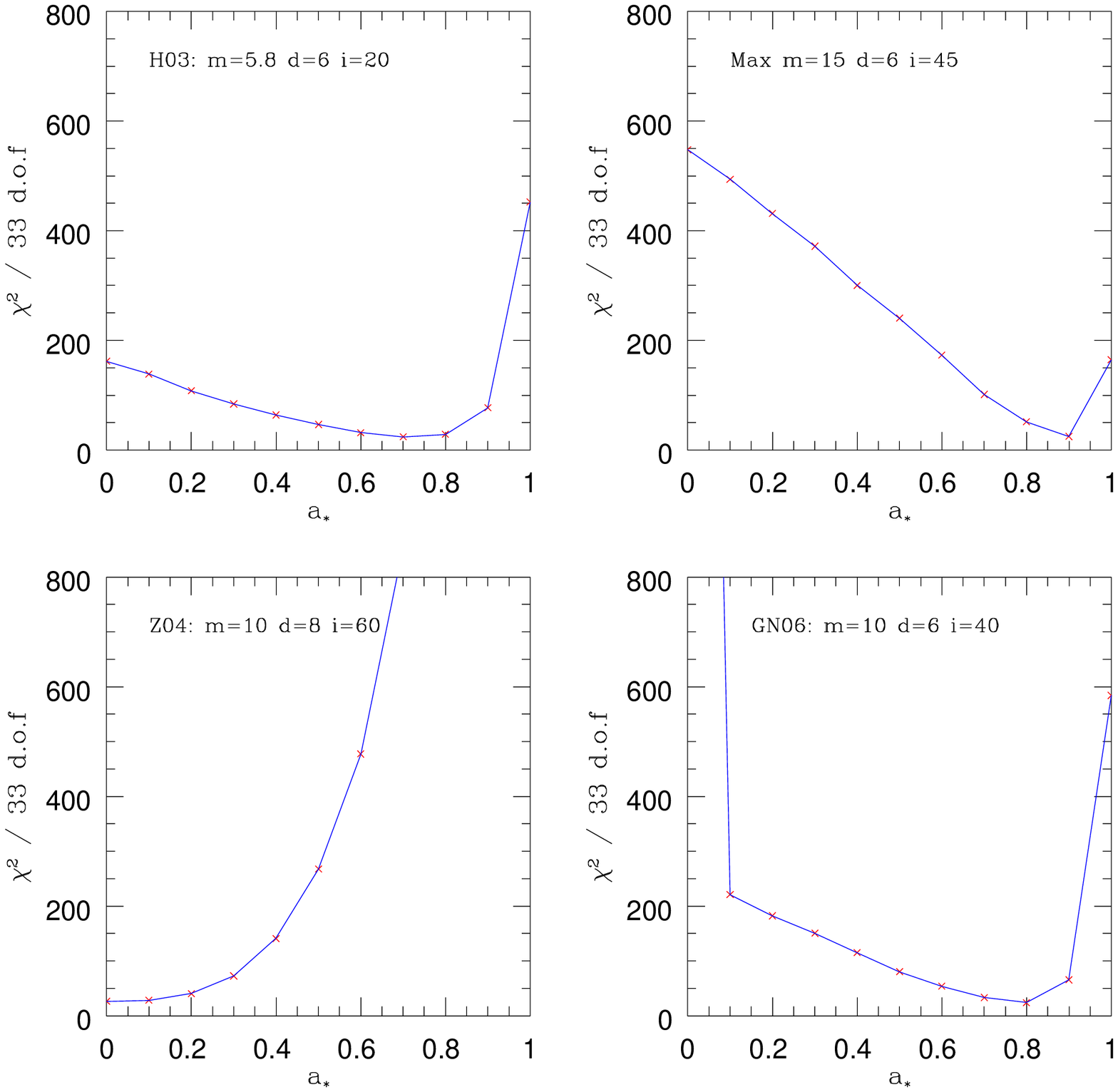} 
\leavevmode\epsfxsize=9cm\epsfbox{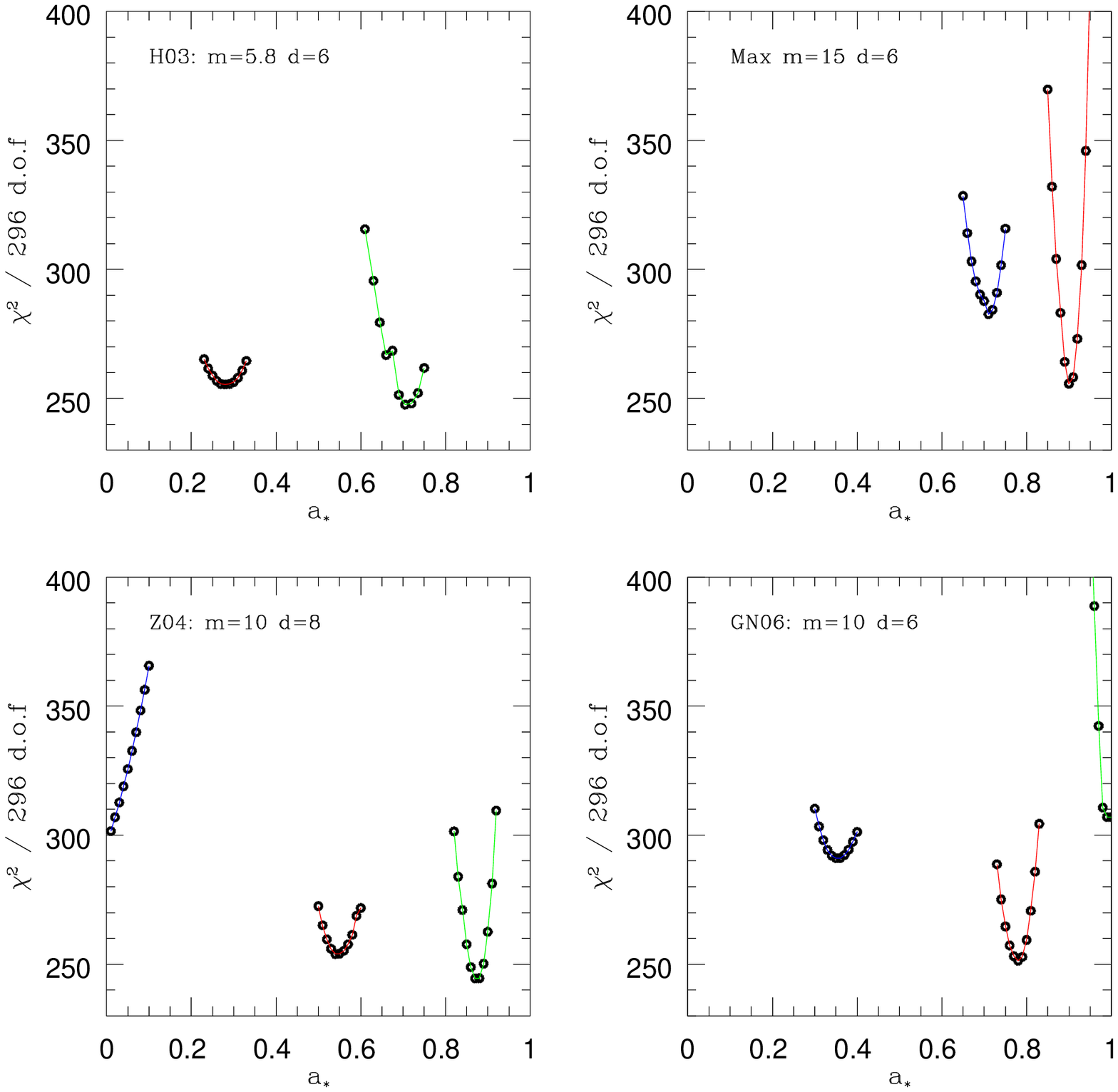} 
\end{tabular}
\caption{{\it Left panels}: The individual spectra best-fit $\chi^{2}$ values with fixed $a_{*}$. {\it Right panels}: The 9 spectra fits for $a_{*}$ with different inclinations. The green colour indicates $i=20^{\circ}$, red $i=40^{\circ}$ and blue $i=60^{\circ}$. $a_{*}$ clearly increases with decreasing inclination.}
\label{9spec}
\end{figure*}

\begin{figure}
\begin{center}
\leavevmode \epsfxsize=8.5cm \epsfbox{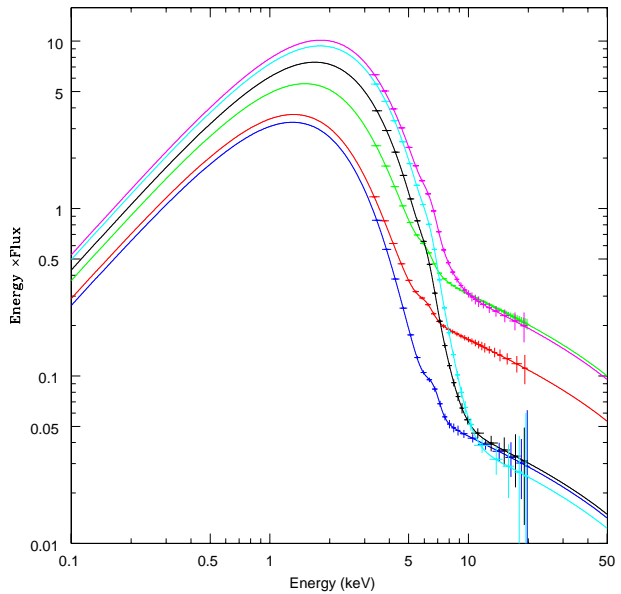} 
\end{center}
\caption{A selection of 6 unfolded spectra plotted with a simple diskbb+thCompml model covering a range of luminosities in our data.}
\label{eemo}
\end{figure}

After fitting all the spectra individually, we chose 9 spectra which
together represent the full scale of the flux (marked by the red
points in Figure~\ref{fluxt}). These 9 spectra were then fitted
together simultaneously with the four different sets of system
parameters. However, this time the spin parameter $a_{*}$ was tied
between the datasets, and fit explicitly.  This allows us to explore
the complex effects of inclination directly, by deriving spin for
three different inclinations, $20^{\circ}$, $40^{\circ}$ and
$60^{\circ}$. The results are plotted in Figure~\ref{9spec} and the
unfolded spectra is plotted with the simple diskbb+thCompml model at
different luminosities in Figure~\ref{eemo}.

The inclinations are illustrated with different colours for easy
comparison; the smallest inclination angle ($20^{\circ}$) is marked
with green, $40^{\circ}$ with red and the largest angle, $60^{\circ}$,
with blue. As is clear from all the fits, increasing inclination angle
decreases the spin. This simultaneous spectral fitting gives the same
results as the multiple individual fits for the same inclination
angle, but gives higher resolution around the spin value e.g. for H03
the minimum appears rather shallow, extending from 0.6--0.8, whereas in
the simultaneous fits it is clear that for $i=20^\circ$ then $a_*=0.7$
is the best fit. For Max, an inclination of $20^\circ$ gives 
$\chi^2>400$ so is not included in the plot. This is because for such
a low inclination, this would imply an emitting area smaller than that
from maximal spin. Similarly, $i=60^\circ$ is not shown on 
H03 as this would imply such a large emission area as to require
retrograde spin. 

\begin{table}
\begin{tabular}{lc|l|l|c}
  \hline
  Parameter set & \textsl{i} &  $a_{*}$ & $\chi^{2}$/ 297 d.o.f
 \\
  \hline
 H03 & $20^{\circ}$ & 0.711$_{-0.013}^{+0.017}$ & 246.62\\
 M= 5.8 $M_{\odot}$ & $40^{\circ}$  & 0.282$_{-0.027}^{+0.028}$ & 255.43\\
 D= 6 kpc  & $60^{\circ}$ & 0.000 (hard limit) & 1010.74  \\
 \hline
 Z04 & $20^{\circ}$ & 0.874$_{-0.010}^{+0.010}$ & 243.75\\
 M= 10 $M_{\odot}$ & $40^{\circ}$ & 0.540$_{-0.012}^{+0.026}$ & 253.85 \\
 D= 8 kpc   & $60^{\circ}$ & 0.000$_{-0.000}^{+0.006}$ & 297.10 \\
 \hline
 GN06 & $20^{\circ}$ & 0.985$_{-0.002}^{+0.014}$ & 302.49 \\
  M= 10 $M_{\odot}$& $40^{\circ}$ & 0.780$_{-0.013}^{+0.015}$ & 251.35\\
  D= 6 kpc  & $60^{\circ}$ & 0.355$_{-0.023}^{+0.024}$ & 291.23\\
\hline
  Max  &	$20^{\circ}$ & 0.998 (hard limit) & 37480.59 \\
  M= 15 $M_{\odot}$& $40^{\circ}$ & 0.946$_{-0.004}^{+0.005}$ & 244.15\\
  D= 6 kpc & $60^{\circ}$ & 0.710$_{-0.010}^{+0.009}$ &  287.64 \\
\hline
\end{tabular}
\caption{The results for the simultaneous spectral fitting showing the best-fit $a_{*}$ values with different inclinations.}
\end{table}

Assuming the inner disc inclination is the same as the orbital
inclination, then the lower limit of $\sim 45^\circ$ implies that
$a_{*}<0.9$ for any reasonable mass ($<15 M_\odot$) and distance
($>6$~kpc). Any lower mass and/or larger distance and/or higher
inclination will give lower spin.

\section{Discussion}

\subsection{Spin from disc spectral fitting} 

There is a hard upper limit on spin of $a_*<0.9$ from the disc spectral
fitting, assuming that the inner disc inclination is the same as that
of the orbit. Our upper limit is very conservative as it is quite
unlikely that the system parameters are all at their extreme values,
and in fact a distance of 6~kpc is inconsistent with a 15~$M_\odot$
black hole.  At 6~kpc the companion star must be towards the 'minimum
mass' solution of Mu\~noz-Darias et al. (2008) in order to get below
the observed r-band magnitude limit of 21.4 (Zdziarski et al. 2004;
corrected from Shahbaz et al. 2001). Such a star has a low mass
transfer rate, yet a large black hole requires a high mass transfer
rate to keep the disc close to the borderline between quiescent and
transient in order to trigger the multiple outbursts (Mu\~noz-Darias
et al. 2008). 

\subsection{Spin from the iron line fits}

There are three independent data sets for GX 339$-$4 where the line
profile has been modelled in detail (Miller et al. 2004; 2006; 2008;
2009; Reis et al. 2008; 2009). We take the most recent determinations,
as these use the best current reflection models for black hole binary
discs (Ross \& Fabian 2007).  The low/hard state XMM-Newton MOS data
give an inner radius of $r_{in}=2.04_{-0.02}^{+0.07}$, inclination
$i=20.0^{\circ}_{-1.3}$ (no upper limit given) with emissivity $3.15\pm 0.15$
(Reis et al. 2008). Suzaku observations of an intermediate state give
$r_{in}=2.39_{-0.29}^{+0.24}$ (i.e. spin of $0.89\pm 0.04$),
inclination $i=18^{\circ}\pm 1^{\circ}$ (no upper limit given) with emissivity
$3.0\pm 0.1$ (Miller et al. 2008). Very high state XMM-Newton burst
mode PN data give a consistent inner radius and inclination of
$r_{in}=2.02_{-0.06}^{+0.02}$ and $i=20.0^{\circ}_{-0.3}$ (no upper limit
given). However, this requires a broken power law emissivity which changes
from $3$ to $7.05_{-0.2}^{+0.05}$ at $r=6$ (Reis et al. 2008).

However, two of these datasets have been challenged as being affected
by instrumental pileup. This 
clearly affects the line determined from the XMM-Newton MOS low/hard
state data, as the simultaneous PN timing mode (which can handle much
higher count rates without pileup) data show a much narrower line
(Done \& Diaz Trigo 2009). Similarly the intermediate state {\em
Suzaku} observation may also be affected by pileup (Yamada et
al. 2009).  However, the PN burst mode very high state data are not
affected by this. These give an equivalent spin of
$0.942^{+0.005}_{-0.004}$ for $i=18^{\circ}\pm 1^{\circ}$ and central emissivity of
$\sim 7$.

\subsection{Comparison of spin from disc fitting and Fe line profile}
\label{s:compare}

Thus there is a clear mismatch between the parameters derived from the
iron line and those derived from disc spectral fitting in GX 339$-$4,
even after excluding piled up data. The only possible way to make the
two consistent are if the inner disc is seen at low inclination, so
that it is misaligned from the binary orbit. This would also make it
consistent with the low inclination of $18^{\circ}\pm 1^{\circ}$ derived from the iron
line fits. However, such a large misalignment could only come about
from a very asymmetric supernovae, but the consequent natal kick is
most likely to unbind the black hole from its binary companion,
disrupting the system entirely (Fragos et al. 2010). This high spin is
also inconsistent with the natal spins predicted from supernovae
collapse models (Gammie et al. 2004), though these are poorly
understood. While accretion does act to spin up the black hole, a low
mass companion star of $\sim 1~M_\odot$ has insufficient mass to
significantly increase the spin of a $\sim 10~M_\odot$ black hole
(King \& Kolb 1999). We note that none of the spins derived from 
disc spectral fitting in low mass X-ray binaries are higher than $0.9$
except perhaps 
GRS 1915+105, with $a=0.98$ claimed by McClintock et al. (2006). However, 
this depends on details of model assumptions 
for the Comptonised spectrum, and can be as low as $a_*\sim 0.7$ (Middleton et al. 2006).
It seems premature to use contested results from this 
pathological source to challenge the supernovae collapse models. 

There are three additional sources where this comparison can be made.
These are 4U~1543-475 (iron line: $a_*=0.3^{+0.2}_{-0.1}$,
$22^\circ$ where {\sc bhspec} gives $a_*=0.75-0.85$) XTE~J1550-564
(iron line: $a_*=0.78\pm 0.02$, $50^\circ$, nearest comparison for
{\sc bhspec} is $43^\circ$ with spin $0.72^{+0.15}_{-0.01}$) and
GRO~J1655-40 (iron line: $a_*=0.94\pm 0.03$ at $70^\circ$ whereas {\sc
bhspec} gives $a_*=0.65-0.75$), with ionised reflection fits in Miller
et al. (2009; Table 3) and disc fits in Shafee et al. (2006) and Davis
et al. (2006).

This means that 2/4 objects (GX 339$-$4 in this paper and GRO J1655-40:
Miller et al. 2009) for which this comparison can be performed give a
significantly larger black hole spin from iron line fitting than from
the disc spectral method.  The remaining 2 objects are marginally
consistent, though taking the uncertainties at face value means that
formally the spins from 4U~1543-475 are also marginally inconsistent
but in the opposite sense (disc fits give higher spin than the iron
line).

Which method (if any!) should we trust? The disc spectrum is the
dominant spectral component, and the derived disc parameters follow
the predicted behaviour for a disc i.e. constant inner radius for
changing mass accretion rate (Ebisawa et al. 1993; Kubota et al. 2001;
Gierli\'nski \& Done 2004). By contrast, the iron line fits are to a
small feature on the total spectrum (which may have a much more
complex form than the typical fit of disc plus power law and its
reflection) and often require a highly centrally concentrated line
emissivity which is not consistent with the simplest expectations of
disc illumination. Thus we argue that the disc spectral fitting model
results are more likely to be robust.

\section{Conclusions}

We derive a hard upper limit for the spin of GX 339$-$4 of $a_*<0.9$
assuming that the inner disc inclination is the same as that of the
binary orbit ($70^\circ<i<45^\circ$). This is inconsistent with the
spin of $0.942^{+0.005}_{-0.004}$ and inclination of 
$i=18^{\circ}\pm 1^{\circ}$ derived from the (non-piled up) XMM-Newton burst mode
very high state data (Reis et al. 2008). This high spin/low inclination
derived from the iron line is already uncomfortably extreme compared to the
lower spins predicted from supernovae collapse models, and the 
small misalignment angles between black hole spin and binary orbit
predicted from binary formation models. While these are both potentially
poorly understood, they are independent constraints and the iron line
profile in GX 339$-$4 conflicts with both of them. 

The iron line profile itself is in subtle conflict with the X-ray
continuum as the reflection smearing parameters require that the
illumination pattern is highly centrally concentrated (Miller et
al. 2008; Reis et al. 2008). Yet the very high state spectral shape
requires that much of the inner disc is covered by optically thick
Comptonizing material, making it very difficult to see strong
reflection from this material (Done \& Kubota 2006).

Thus we argue that the spin/inclination/emissivity derived from the
iron line profile are uncomfortably extreme. It seems far more likely
to us that the more moderate spin implied by the disc spectral fitting
results where the inner disc is more or less aligned with the binary
orbit are giving a more robust answer. The inescapable corollary to
this is that there are systematic effects affecting spin as derived
from the iron line profile that are not yet understood. This is a
crucial issue in applying the iron line models with confidence to
derive spin in AGNs.

\label{lastpage}

\end{document}